\documentclass[12pt]{iopart}
\usepackage{iopams}
\usepackage{epsf}
\begin{document}

\title{Coincidence analysis to search for inspiraling compact binaries}

\author{Hirotaka Takahashi\dag \ddag 
\footnote[3]{E-mail:hirotaka@vega.ess.sci.osaka-u.ac.jp},
Hideyuki Tagoshi\dag\, the TAMA Collaboration and the LISM Collaboration
}

\address{\dag\ Department of Earth and Space Science , 
Graduate School of Science, Osaka University, 
Toyonaka, Osaka 560--0043, Japan}

\address{\ddag\ Graduate School of Science and Technology, 
Niigata University, Niigata, Niigata 950-2181, Japan}

\begin{abstract}
We discuss a method of coincidence analysis to search 
for gravitational waves from inspiraling compact binaries
using the data of two laser interferometer gravitational wave detectors. 
We examine the allowed difference of the wave's parameters
estimated by each detector to obtain good detection efficiency. 
We also discuss a method to set an upper limit to the event rate
from the results of the coincidence analysis. 
For the purpose to test above methods, 
we performed a coincidence analysis by applying these methods 
to the real data of TAMA300 and LISM detectors 
taken during 2001. 
We show that the fake event rate is reduced significantly by 
the coincidence analysis without losing real events very much. 
Results of the test analysis are also given. 
\end{abstract}




\section{Introduction}
Several laser interferometric gravitational wave detectors, such like 
TAMA300[1], LIGO[2], GEO600[3], and VIRGO[4], have already been constructed or
expected to be finished its construction soon. 

TAMA300 is an interferometric 
gravitational wave  detector with $300$ m baseline length located at 
Mitaka campus of the National Astronomical Observatory of Japan in Tokyo 
$(35.68 ^{\circ}N,139.54 ^{\circ}E)$.
LISM is an interferometric gravitational wave detector 
with $20$ m baseline length located at Kamioka mine, Gifu 
$(36.25 ^{\circ}N,137.18 ^{\circ}E)$.
LISM was originally developed as a prototype detector
during 1990 to 1995 
in the National Astronomical Observatory in Mitaka, Tokyo. 
In 2000, it was moved into the Kamioka mine to perform observation. 

TAMA300 began to operate in 1999, and performed 
observations for 7 times by the end of 2002. 
In particular, during the period from 1 August to 20 September 2001, TAMA300 
performed the longest observation, and about 1100 hours of data were taken. 
LISM detector also performed an observation from  1 to 23 August, 
and 3 to 17 September, 2001, and about 780 hours of data were taken. 
This observation is called Data Taking 6 (DT6) 
among the TAMA collaboration and the LISM collaboration.
Among the LISM data, the last half of data taken during September are
in good condition, 
and are available for the gravitational wave event searches. 
As a result, the length of data available for the coincidence analysis 
is about 275 hours in total. 
The best sensitivity of the TAMA300 was about
$5\times10^{-21}/\sqrt{\rm Hz}$ around $800$Hz. 
The best sensitivity of the LISM 
was about $6.5\times10^{-20}/\sqrt{\rm Hz}$ around $800$Hz. 

In the past, only a few works have been done for 
the coincidence analysis using real data of two or more laser interferometers, 
although many works has been done using bar detectors. 
As far as we are aware of, only one coincidence analysis 
to search for burst events 
in a pair of laser interferometers 
has been reported \cite{Nicholson}. 
Although, the sensitivities of TAMA300 and LISM are different for 
one order of magnitude, it is a very good opportunity to perform
coincidence analysis since long data are available, and 
both detector have shown good stability which 
allow us to perform such analysis. 

We consider gravitational waves from inspiraling compact binaries, consisting of neutron 
stars or black holes. 
Since their wave forms are well known by post-Newtonian approximation 
of general relativity, we can use the matched filtering. 
In this paper, 
we discuss a method of coincidence analysis to search for inspiraling
compact binaries by using the results of the matched filtering search
by several interferometers. 

The outline of the analysis discussed here is as follows. 
We independently performed a matched filtering search in each detector 
and obtain event lists. 
We compare the lists to find coincident events.
In matched filtering search, each event depends not 
only on the time of coalescence but also on two masses, and so on. 
We can thus require consistency conditions for such parameters 
by which we decide whether each pair of events 
in the event lists can be considered as a coincident event. 
By requiring such condition, we can reduce the false alarm rate significantly. 
It is important to determine the consistency condition so that we do not
lose real events very much. 
We evaluate the errors of estimated parameters by using Fisher matrix 
and by monte carlo simulations. Those estimate of errors are used 
in the coincident event search 
as windows of parameters allowed for each pair of events. 
In real data analysis, it is important to check the detection efficiency. 
Thus, we briefly discuss 
the detection efficiency by using test signals injected into 
TAMA300 and LISM data. 

To test the performance of the above method, we perform the coincidence 
analysis using short length of data of TAMA300 and LISM. 
We find that significant number of fake events are removed by coincidence
analysis without losing real signals significantly. 

After comparing the event lists by requiring consistency conditions,
we have a list of candidate events. In order to state any statistical 
significance of these events, we need to compare the number of survived events
with estimated number of coincident events produced accidentally by fake events. 
We discuss how to estimate the number of events produced accidentally. 

Finally, we discuss a method to set an upper limit to the event rate using
coincident event lists. To test the method, this method is applied to 
the above test results. 
The results for the upper limit given here can not be considered as 
our final scientific results, since our objective here is to test the 
method using the short length of real interferometers' data. 
Complete results of the analysis using all the data of TAMA300 and LISM 
during 2001 will be discussed in a separated paper \cite{takahashi}. 

This paper is organized as follows. In section 2, 
theoretical wave forms, 
basic formulas of matched filtering,
and $\chi^2$ method to reject non-stationary, non-Gaussian
noise events are shown. 
In section 3, a method of coincidence analysis 
are proposed, and detection efficiency are evaluated using TAMA300 and LISM data. 
In section 4, We apply the above method to real data of TAMA300 and LISM, 
and obtain a coincident event list. Using the list, we evaluate the number of
accidental coincident events, and discuss the statistical significance of events.
In section 5, we discuss a method to set an upper limit to the Galactic events. 
Section 6 is devoted to the summary and discussions. 

\section{Matched filtering}
We assume that the time sequential data of the detector output $s(t)$ 
consists of a signal plus noise $n(t)$. 
To characterize the detector's noise, we denote the one-sided power spectrum density
of noise by $S_n(f)$. 
We also assume that the wave forms of
the signals are predicted theoretically with sufficiently good accuracy. 
We call these wave forms as {\it templates}.
We adopt templates calculated by using the post-Newtonian approximation of
general relativity\cite{Blanchet}. We use the restricted  
post-Newtonian wave forms in which the phase evolution is calculated to 2.5 
post-Newtonian order, but the amplitude evolution contains only the lowest 
Newtonian quadrupole contribution. We also use the stationary phase 
approximation to calculate the Fourier transformation of the wave forms.

We denote the parameters distinguishing different templates 
by $\theta^{\mu}$. 
They consist of the coalescence time 
$t_c$, the chirp mass $\mathcal{M} (\equiv M\eta^{3/5})$ 
$(M =m_1+m_2)$, and non-dimensional reduced mass 
$\eta (\equiv m_1 m_2 /M^2)$. 
In this analysis, we did not take into 
account of the effects of spin angular momentum. 
The templates corresponding to a given set of $\theta^{\mu}$ 
are represented in Fourier space by 
two independent templates $\tilde{h}_c$ 
and $\tilde{h}_s$ as 
\begin{equation}
\tilde{h}(f)=\tilde{h}_c(f) \cos\phi_c+\tilde{h}_s(f) \sin\phi_c,
\end{equation}
where $\phi_c$ is the phase of wave, and 
\begin{eqnarray}
\tilde h_c(\theta^{\mu},f) &= &
i \tilde h_s(\theta^{\mu},f)=
{\cal N} f^{-7/6} e^{i(\psi_{\alpha}(f)+t_c f)},\cr
&&\quad \mbox{for}~~ 0<f\leq f_{max}(\theta^{\mu}),\cr
 \tilde h_c(\theta^{\mu},f) & =&\tilde h_s(\theta^{\mu},f)=0,\cr
&&\quad \mbox{for}~~ f> f_{max}(\theta^{\mu}).
\end{eqnarray}
Here ${\cal N}$ is a normalization constant, and
\begin{equation}
\psi_{\alpha}(f)=\sum_i \alpha^i(\theta^{\mu})\zeta_i(f),
\end{equation}
with
\begin{eqnarray}
 &&\alpha^1={3\over 128\eta}(\pi M)^{-5/3},
\cr
 &&\alpha^2={1\over 384\eta}\left({3715\over 84}+55\eta\right)
  (\pi M)^{-1},
\cr
 &&\alpha^3=-{3\pi\over 8\eta}(\pi M)^{-2/3},
\cr
 &&\alpha^4={3\over 128\eta}\Bigl({15293365\over 508032}
   +{27145\over 504}\eta+{3085\over 72}\eta^2\Bigr) (\pi M)^{-1/3},
\cr
 &&\alpha^5={\pi\over 128\eta}
   \left({38645\over 252}+5\eta \right),
\cr
 &&\zeta_1(f)=f^{-5/3},\quad
 \zeta_2(f)=f^{-1},\quad
 \zeta_3(f)=f^{-2/3},\cr
&& \zeta_4(f)=f^{-1/3},\quad
 \zeta_5(f)=\ln f.
\label{eq:temppara}
\end{eqnarray}
Negative frequency components are given by the reality 
condition of $h(t)$ as $\tilde h(-f)={\tilde h^*(f)}$
where $*$ means the operation of taking the complex conjugate.
The value of $f_{max}$ is determined by 
a cut off frequency chosen close to the innermost stable circular 
orbit frequency, $f_c$, or the half of the sampling frequency of data, $f_{s}$
We adopt smaller one from $f_c$ and $f_s$ as $f_{\rm max}$. 

We define a filtered output by
\begin{equation}
\rho (t_c,m_1,m_2,\phi_c) \equiv 
2 \int _{-\infty} ^{\infty} \frac{\tilde{s} (f) \tilde{h}^* (f)}{S_h (f)} df 
= (s|h). \label{eqn:rho} 
\end{equation}
In eq.
(\ref{eqn:rho}), we can analytically take the maximization over $\phi_c$
which gives
\begin{equation}
\rho(t_c,m_1,m_2)
= \sqrt{(s|h_c) ^2 +(s|h_s) ^2} .
\end{equation}

We choose the normalization constant ${\cal N}$ of the templates 
so that it satisfies $(h_c|h_c)=$ $(h_s|h_s)=1$. 
We can see that 
$\rho$ has an expectation value $\sqrt{2}$ in the presence of only Gaussian 
noise. Thus, the signal-to-noise ratio is given by $SNR=\rho/\sqrt{2}$ 

Analyzing the real data of TAMA300, 
we have found that the noise contained a large amount of 
non-stationary and non-Gaussian noise \cite{tama1}.
In order to remove the influence of such noise,
we introduce a $\chi^2$ test of the time-frequency behavior of the signal
\cite{Allen}.
Here, $\chi^2$ is defined as follow.
First we divide each template into $n$ mutually independent pieces
in the frequency domain,
chosen so that the expected contribution to $\rho$ from 
each frequency band is approximately equal, as 
\begin{equation}
\tilde{h}_{c,s}(f) = \tilde{h}_{c,s}^{(1)}(f)+ \tilde{h}_{c,s}^{(2)}(f)
+\cdots + \tilde{h}_{c,s}^{(n)}(f). 
\end{equation}
We calculate
\begin{eqnarray}
z_{(c,s)}^{(i)}=(\tilde{s}|\tilde{h}_{c,s}^{(i)}) ,\nonumber \\
\bar{z}_{(c,s)}^{(i)}=
{1\over n}
(\tilde{s}|\tilde{h}_{c,s}),
\end{eqnarray}
The $\chi^2$ is defined by
\begin{equation}
\chi^2 = n \sum_{i=1}^{n} \Big[{(z_{(c)}^{(i)}-\bar{z}_{(c)}^{(i)})^2
+(z_{(s)}^{(i)}-\bar{z}_{(s)}^{(i)})^2}\Big] .
\end{equation}
This quantity must satisfy the $\chi^2$-statistics with
$2n-2$ degrees of freedom, 
as long as the data consists of Gaussian noise plus 
chirp signals. 
For convenience, we renormalized $\chi^2$ as 
$\chi^2/(2n-2)$. In this paper, we chose $n=16$. 

The value of $\chi^2$ is independent to the amplitude of the signal as long as 
the template and the signal have an identical wave form. However, in reality, 
since the template and the signal have different value of parameters because of
the discrete time step and discrete mass parameters we search,  
the value of $\chi^2$ becomes larger when the amplitude
of signal becomes larger. In such situation, if we reject events simply by 
the value of $\chi^2$, we may lose real events with large amplitude. 
We thus introduce a statistic, 
$\rho/\sqrt{\chi^2}$, to distinguish between candidate events
and noise events.
By checking the Galactic event efficiency, we found that the selection 
by the value of $\rho/\sqrt{\chi^2}$ gives reasonable detection efficiency.

We searched for the mass parameters, 
$1.0 M_{\odot} \le m_1,m_2 \le 2.0 M_{\odot}$, 
which is a typical mass region of neutron stars. In the mass parameter 
space, we prepared a mesh. 
The mesh points define templates used for search.
The mesh separation is determined so that the maximum loss of SNR 
becomes less than $3\%$. We use a special parametrization of mass parameters 
which was introduced by Tanaka and Tagoshi\cite{tanaka-tagoshi}.
This method simplifies algorithm using geometrical arguments to determine the 
mesh points.
The parameter space defined in our search program turned out
to contain about $200 \sim 1000$ templates for the TAMA300 data, and about
$200 \sim 600$ templates for the LISM data.
The variation of the number of template is due to the variation of the shape 
of noise power spectrum. The typical value of the number of template is about
700 for TAMA300, and 400 for LISM.

We perform matched filtering search using 
TAMA300 and LISM data independently. 
We obtain $\rho$ and $\chi^2$ as functions of masses
and the coalescing time $t_c$. In each small interval of coalescing time 
$\Delta t_c$, we looked for an event which had the maximum $\rho$.
In the search we report in the following sections, 
we choose $\Delta t_c = 25$msec. 

\section{A coincidence analysis}
\subsection{Algorithm}
Each event in the event list, obtained 
by the matched filtering search independently performed 
for two detectors, depends on $t_c$, $M$, and $\eta$. 
If they are real
events, they should have the same 
parameters in 
both event lists. 
However, we may observe real events with different parameters by 
the effects of detectors' noise and other effects. 
Therefore we have to determine the allowed difference of parameters
by taking 
into account of these effects, in order not to lose real events by coincidence
analysis. 

{\it Time selection}: 
First we discuss the coalescence time $t_c$. 
The distance between TAMA300 and LISM is 
$219.92$km.
Therefore, the
maximum delay of the signal arrival time is 
$\Delta t_{\rm dis}= 0.73$msec.
The effect of detectors' noise to the estimated value of $t_c$ can be 
evaluated by the 
Fisher information matrix\cite{Cutler}.  
We denote the $1\sigma$ value of error for each detector 
as $\Delta t_{c \mbox{\rm\tiny TAMA}, \mbox{\rm\tiny LISM}}$. 
We can determine allowed error of $t_c$ due to noise by 
$\Delta t_{\rm noise}=\sigma \times \Delta t_{c}$ where 
$\Delta t_{c} \equiv \sqrt{\Delta t_{c \mbox{\rm\tiny TAMA} }^2
+\Delta t_{c \mbox{\rm\tiny LISM}}^2}$.

Finally, we define the allowed difference of $t_c$ as follows. 
If the parameters $t_c^{\mbox{\rm\tiny TAMA}},\ t_c^{\mbox{\rm\tiny LISM}}$ of the each pair of events satisfy 
\begin{equation}
|t_c^{\mbox{\rm\tiny TAMA}}-t_c^{\mbox{\rm\tiny LISM}}|< \Delta t_{\rm dis}+\Delta t_{\rm noise} ,
\end{equation}
the event is recorded as a candidate event. 

Note that the Gaussian distribution of the noise and 
the large amplitude of the events are assumed in evaluating
the $1\sigma$ error of the estimated parameter due to noise. 
Thus, in applying the real data analysis in which the noise is not necessary 
Gaussian, and the amplitude of events are not necessary very large, 
we have to check the detection efficiency by other methods. 
As discussed in the next section, we find that 
if we adopt $\sigma>3$, we will be able to obtain very high 
detection probability even in the case of real data.

{\it Mass selection}:
Next we discuss the mass parameters. 
The error of estimated parameter of the chirp mass and reduced mass 
due to noise can also be evaluated by the Fisher matrix. 
We denote it by $\Delta \mathcal{M}_{\rm noise}$ and $\Delta \eta_{\rm noise}$,
which are evaluated from each detector as 
\begin{eqnarray}
\Delta \mathcal{M}_{\rm noise}
=\sigma \sqrt{\Delta \mathcal{M}^2_{\mbox{\rm\tiny TAMA}}+
\Delta \mathcal{M}^2_{\mbox{\rm\tiny LISM}}}, \nonumber\\
\Delta \eta_{\rm noise}=\sigma \sqrt{\Delta \eta^2_{\mbox{\rm\tiny TAMA}}+
\Delta \eta^2_{\mbox{\rm\tiny LISM}}}, \nonumber
\end{eqnarray}
where $\Delta \mathcal{M}_{i}$ and $\Delta \eta_i$ are $1 \sigma$ value of error
induced by each detector's noise. 
The value of $\sigma$ will be discussed in the next subsection. 

Along with the error due to noise, 
we also have to take into account of the effect of the finite mesh size.
When the amplitude of the signal is very large, the errors evaluated by 
the Fisher matrix becomes smaller than the value of finite mesh size,
since the error of the parameter due to noise is inversely proportional to 
the value of $\rho$. 
We denote the error due to the finite mesh size as 
$\Delta \mathcal{M}_{mesh}$, $\Delta \eta_{mesh}$. 
We determine the allowed difference of the chirp mass and the reduced mass 
so that 
if the parameters $\mathcal{M} ^{\mbox{\rm\tiny TAMA}},
\ \mathcal{M} ^{\mbox{\rm\tiny LISM}}$,
$\eta^{\mbox{\rm\tiny TAMA}}$ and $\eta^{\mbox{\rm\tiny LISM}}$ 
of each pair of events satisfy
\begin{eqnarray}
|\mathcal{M} ^{\mbox{\rm\tiny TAMA}} - \mathcal{M} ^{\mbox{\rm\tiny LISM}}| &<& 
\max(\Delta \mathcal{M}_{\rm noise},\Delta \mathcal{M}_{\rm mesh}),\nonumber \\
|\eta ^{\mbox{\rm\tiny TAMA}} - \eta ^{\mbox{\rm\tiny LISM}}| &<& 
\max(\Delta \eta_{\rm noise},\Delta \eta_{\rm mesh}) ,
\end{eqnarray}
the pair of event is adopted as a candidate event. 

{\it Amplitude selection}:
Next we discuss the amplitude. 
When the sensitivity of the detectors is different, 
the signal-to-noise ratio observed by each detector will be different. 
Further, since the direction of the arm of each interferometer will be different
in general, 
the signal-to-noise ratio will be different in each detector
even if the sensitivity is the same. Even in such cases, we can still 
require consistency condition to the events and reduce the number of fake events.

In our case, the sensitivity of TAMA300 is typically 15 times better than 
that of LISM. 
We can simply express this effect by $\delta_{\rm sens}$ defined as
\begin{equation}
\delta _{\rm sens}\equiv 
\log \Big[\Big(\int \frac{f^{-7/3}}{S_{n\ \mbox{\rm\tiny TAMA}}(f)} df \Big)^{1/2}/
\Big(\int \frac{f^{-7/3}}{S_{n\ \mbox{\rm\tiny LISM}}(f)} df \Big)^{1/2}\ \Big] .
\end{equation}
The arm direction of LISM is rotated from that of TAMA300 for 60 degrees.
Real events will be detected with different SNR by each 
detector depending on its incident direction and polarization. 
A simple and straightforward way to evaluate 
the allowed difference of amplitude is to perform simulations. 
To evaluate only the effect of different arm direction, 
we assume two detectors have identical noise power spectrum. 
We them perform simulations by generating the Galactic events 
and by evaluating the difference of $\rho$ which are detected by 
each detector. We them determine the value of $\delta_{\rm simu}$ 
such that for more than 99.9 \% of events, we have 
\begin{equation}
-\delta_{\rm simu}\leq \log\left({\frac{\rho_{\mbox{\rm\tiny TAMA}}}
{\rho_{\mbox{\rm\tiny LISM}}}}\right)
\leq \delta_{\rm simu}. 
\end{equation}
There is also an error due to noise in the estimated value of $\rho$ which
can also be estimated by the Fisher matrix in the same way as $t_c$, $\mathcal{M}$
and $\eta$. We denote it as $\delta_{\rm noise}$. 

By combining the above two effects, 
we determine the allowed difference of $\rho_{TAMA}$ and $\rho_{LISM}$ by
\begin{equation}
\delta _{\rm sens} - \delta _{\rm simu} -\delta_{\rm noise}\le 
\log \Big({\frac{\rho_{\mbox{\rm\tiny TAMA}}}{\rho_{\mbox{\rm\tiny LISM}}}}\Big) 
\le \delta _{\rm sens} + \delta _{\rm simu}+\delta_{\rm noise}. 
\end{equation}

\subsection{Efficiency}
\label{sec:efficiency}
Here, we discuss the detection efficiency of the algorithm of the coincidence analysis 
discussed in the previous subsection. 
As model sources, we generate inspiraling compact binaries located within 1 kpc 
from Solar system in our Galaxy. The distribution of the binaries are 
determined so that it reproduce a density distribution \cite{Allen}.
Other parameters like two masses, the inclination angle, the polarization angle, and 
phase are determined randomly. We consider the mass 
from $1M_\odot$ to $2M_\odot$ for each star. 
We inject 1000 events into TAMA300 and LISM data, and analyse the data using
the algorithm discussed in the previous section. 
When we set a threshold $\rho>7$, the detection efficiently of 
the independent analysis of TAMA300 and LISM 
becomes 99\% and 33\% respectively. 
Since we perform coincidence analysis, the maximum value of the detection efficiency 
is limited by LISM's efficiency.  
We thus show the relative efficiency of the analysis, after applying 
criteria of coincidence,  
compared to the one detector efficiency of LISM in Fig.1. 
We find that, among 33 \% of events which are detected by LISM, 
we can detect even after the coincidence analysis more than 94 \% 
if we set $\sigma>3$. 
Thus, in the analysis discussed in the next section, we adopt a precise 
$\sigma=3.29$ because  
it corresponds to $0.1\%$ probability of 
losing real events in Gaussian noise case.

\section{Application to TAMA300 and LISM data}
\subsection{Results of coincidence analysis}
We perform a matched filtering search using TAMA300 and LISM data 
respectively. 
For the purpose of test analysis, 
we use 26.0 hours of data for which both detector were locked. 
This length is about 10$\%$ of all the data available for coincidence
analysis taken during Data Taking 6. 
As results of matched filtering search, 
there are 159,935 events for TAMA300 and 109,609  events 
for LISM. 

For these candidate events, we perform a coincidence 
analysis by requiring consistency among the parameters. 
The consistency conditions are imposed 
in the order of the coalescence time selection, the mass selection 
and the amplitude selection. 
In Table \ref{tab:coincident}, we show the results
of the coincident event search.
In Fig.2, we also show a scatter plot of the coincident events 
after coincidence selections in terms of the value 
of $\rho_{\mbox{\rm\tiny TAMA}}$ and 
$\rho_{\mbox{\rm\tiny LISM}}$. 

Although significant number of fake events are removed by taking 
coincidence, there still remains many events with large $\chi^2$. 
Thus, we can obtain better detection efficiency if we introduce a $\chi^2$
selection in addition to the coincidence selections. 
In Fig.3, we also show a scatter plot of the coincident events 
after coincidence selections in terms of the value 
of $\rho_{\mbox{\rm\tiny TAMA}}/\sqrt{\chi^2_{\mbox{\rm\tiny TAMA}}}$ and 
$\rho_{\mbox{\rm\tiny LISM}}/\sqrt{\chi^2_{\mbox{\rm\tiny LISM}}}$. 

We examine the distribution of $\rho$ and $\sqrt{\chi^2}$ for each 
detector. The peak value of 
$\rho_{\mbox{\rm\tiny TAMA}} $,$\rho_{\mbox{\rm\tiny LISM}}$, 
$\sqrt{\chi^2_{\mbox{\rm\tiny TAMA}}}$ and 
$\sqrt{\chi^2_{\mbox{\rm\tiny LISM}}}$
are 8.90, 10.18, 2.38, 2.80. 
It is expected from this that the value of $\rho/\sqrt{\chi^2}$ 
of the coincident events are distributed around $\rho/\sqrt{\chi^2}\sim 4$.
In Fig.3, we find that the distribution of the coincident events 
are consistent with this expectation. 

By injecting test signals into data, it is confirmed that 
when signals with $\rho=20$ are detected, the value of $\sqrt{\chi^2}$
become $\sqrt{\chi^2}\sim 1.3$, Such events have value 
$\rho/\sqrt{\chi^2}\sim 15$. Thus, from Fig.3, 
if such events really happened, 
the value of $\rho/\sqrt{\chi^2}$ would become much larger than the 
tail of the distribution of coincident events. 
For TAMA300, the value $\rho\sim 20$ will be produced by 
events occurring in the center of the Galaxy, 
although the distance will be much shorter for LISM. 


\begin{table}[htpd]
\begin{center}
\caption{Results of the coincidence analysis. 
$n_{\rm obs}$ is the number of coincident events. 
$\bar{n}_{acc}$, $\bar{\sigma}_{acc}$ are  
the estimated number of accidental coincidence
and its standard deviation.  }
\begin{tabular}{c c c }
\hline 
  & $n_{obs}$ & $\bar{n}_{acc} \pm \bar{\sigma}_{acc}$    \\
\hline 
after time selection & 486 & $ 579.13 \pm 248.57 $   \\
after time and mass selection& 74 & $ 84.68 \pm 34.72 $  \\
after time, mass and amplitude selection & 63 & $67.14 \pm 32.90 $ \\
\hline
\end{tabular} 
\end{center} 
\label{tab:coincident} 
\end{table}

\subsection{Accidental coincidence}
\label{sec:acc}
In this section we discuss how to estimate the number of accidental coincident 
events. 
It is possible to estimate the number of accidental coincident events
by a usual procedure of shifting one of two sets of data by a time $\delta t$ 
(called time delay) and determining the number of coincidence 
$n_{c}(\delta t)$ \cite{Amaldi} \cite{Astone}.
Repeating for $m$ different values of time delay, 
the expected number of coincidence is simply the sample mean of the $m$
values of $n_{c}(\delta t)$ at delays other than $\delta t=0$,
\begin{equation}
\bar{n}_{acc} = \frac{1}{m}\sum_{i=1}^{m} n_{c}(\delta t_{i}) . 
\end{equation}
The corresponding experimental standard deviation is given by
\begin{equation}
\bar{\sigma}_{acc} = \sqrt{\sum_{i=1}^{m} \Big( n_{c}(\delta t_{i}) 
- \bar{n}_{acc} \Big ) ^2 / m} . 
\end{equation}
Since there is no signal at delays other than $\delta t=0$, the number of 
coincident events at these delays can be considered as an experimental estimation
of the parent distribution from which the accidental coincidence are drawn.

In this analysis, we performed the time shift for 400 times 
between $-40,000$ seconds and $40,000$ seconds. 
Each time shift is separated with another time shift for $200$ seconds. 

The number of accidental coincidence and its standard deviation is shown 
in Table.1.
We can see that 
the number of coincident events obtained after each selection
is completely agree with the number of accidental coincidence. 
Thus, we conclude that we find no signature of 
gravitational wave events in the data used here. 

By examining the event rate of each detector in each $100$seconds interval,
we find that 
the events rate are nearly stationary in the period of 26 hours 
in both detectors, 
except a few portions which show the event rate $7\sim 10$ larger than
the other normal portions. 
Since we did not excluded such burst like portions in the time shift 
procedure, the average and the variance of the number of accidental 
coincident events are slightly affected by such portions. 
If we exclude such portions, the average of accidental 
coincidence becomes smaller, and the value becomes more closely 
to the observed value. The variance also becomes smaller. 
However, since the main purpose of the analysis here is to test our 
coincidence procedure, we do not discuss it furthermore here. 


\section{Upper limit to the Galactic event rate}
In this section, 
we discuss a method to evaluate the upper limit 
to the Galactic event rate based on the result of the coincidence analysis. 
As sources, we consider nearby events within 1kpc from the Solar system
in our Galaxy. 

The upper limit to the event rate is given by
$N/(T \epsilon)$ ,where $N$ is an upper limit to the average number of 
real events over a certain threshold, $T$ is length of data [hours] and
$\epsilon$ is the detection efficiency. 

First, we evaluate an upper limit to  
average number of real events, $N$, over a threshold. 
Assuming Poisson distribution for the number of real/fake events 
over a threshold, we can obtain an upper limit to the 
expected number of real events, $N$, as a solution of \cite{particle} 
\begin{equation}
\frac{e^{-(N+N_{bg})}\sum_{n=0}^{n=N_{OBS}} \frac{(N+N_{bg})^n}{n!}}
{e^{-(N_{bg})}\sum_{n=0}^{n=N_{OBS}} \frac{(N_{bg})^n}{n!}} = 1 - {\rm C.L.} .
\label{eq:poisson}
\end{equation}
where $N_{obs}$ is an observed number of events over the threshold and $N_{bg}$ is
an estimated number of fake events over the threshold, C.L. is 
a confidence level. 

In order to determined the threshold, 
in stead of defining the single signal-to-noise ratio, 
we define thresholds to the value of 
$\rho_{\mbox{\rm\tiny TAMA}}/\sqrt{\chi_{\mbox{\rm\tiny TAMA}}^2}$ and 
$\rho_{\mbox{\rm\tiny LISM}}/\sqrt{\chi_{\mbox{\rm\tiny LISM}}^2}$. 
{}From Fig.\ref{fig:scatter2}, we set a threshold for 
TAMA300 to $\rho_{\mbox{\rm\tiny TAMA}}/\sqrt{\chi_{\mbox{\rm\tiny TAMA}}^2}=7.27$, and for LISM to
$\rho_{\mbox{\rm\tiny LISM}}/\sqrt{\chi_{\mbox{\rm\tiny LISM}}^2}=7.47$. 
Observed number of events over these thresholds is $N_{\rm obs}=0$.
We can also evaluate 
expected number of fake events over thresholds to be $N_{\rm bg}=0.74$\, 
which corresponds to the false alarm rate,  
0.028 events/hours. 
Using Eq.(\ref{eq:poisson}), we obtain upper limit to the 
average number of real events over the threshold
as $N=2.30\ (C.L.=90\%)$.

Second, we evaluate detection efficiency $\epsilon$ by 
simulations explained in section \ref{sec:efficiency}. 
With the threshold determined above, 
we obtain the probability that we 
detect events over the each detector's threshold as 
$\epsilon= 0.20$.

Using above results and the length of data $T=26.0$ [hours],
we obtain an upper limit to the  
event rate of near by sources in 1kpc to be 
$N/(T \epsilon) = 0.44$ events/hour (C.L.= $90\%$).
This value is not improved from the value obtained by 
the analysis of TAMA300 alone. This is because the sensitivity of TAMA300 
is nearly 100 \% in this distance, although the sensitivity of LISM 
is still less than 30 \%. In such situation, the gain of efficiency 
obtained due to lower threshold is dominated by the lower sensitivity 
of LISM, since the efficiency of coincidence is limited by the detector 
with lower sensitivity. 

\section{Summary}
In this paper, 
we discussed a method of the coincidence analysis 
to search for inspiraling compact binaries using matched filtering. 
We examined the selection criteria for coincidence of the parameters 
like the time of coalescence, chirp mass, reduced mass, 
and the amplitude of events. 
We found, by simulation using the data of TAMA300 and LISM, that
we do not lose events significantly by appropriately
choosing the selection criteria. 
The method discussed in this paper
can easily applicable to multiple detectors cases in which more than 
two detectors are used. 

For the test purpose, we performed coincidence analysis using 
the short length of data of 
TAMA300 and LISM taken during DT6 observation in 2001. 
We found that significant number of fake events are removed by taking
coincidence. 

Since the sensitivity of TAMA300 is much better than LISM,
the detection efficiency after coincidence analysis does not improved 
compared to that of TAMA300 alone, although we can have a better 
efficiency compared to that of LISM alone. 
Thus, we do not have an improved upper limit to the event rate compared to 
the result obtained by TAMA300 alone. 
However, when two detectors have comparable sensitivity, 
it is possible to obtain an improved sensitivity
compared to the one detector analysis. 
Further, in any cases, when we find candidate gravitational wave events,
its statistical significance becomes larger by taking coincidence. 

Complete results of the analysis, including 
an upper limit to the event rate using all the data of TAMA300 and LISM 
during Data Taking 6 will be reported elsewhere \cite{takahashi}. 

This work was supported in part by Grant-in-Aid for Scientific Research,
Nos.~14047214 and 12640269, of the Ministry of Education, 
Culture, Sports, Science, and Technology of Japan.

\begin{figure}[htpd]
\begin{center}
\epsfbox{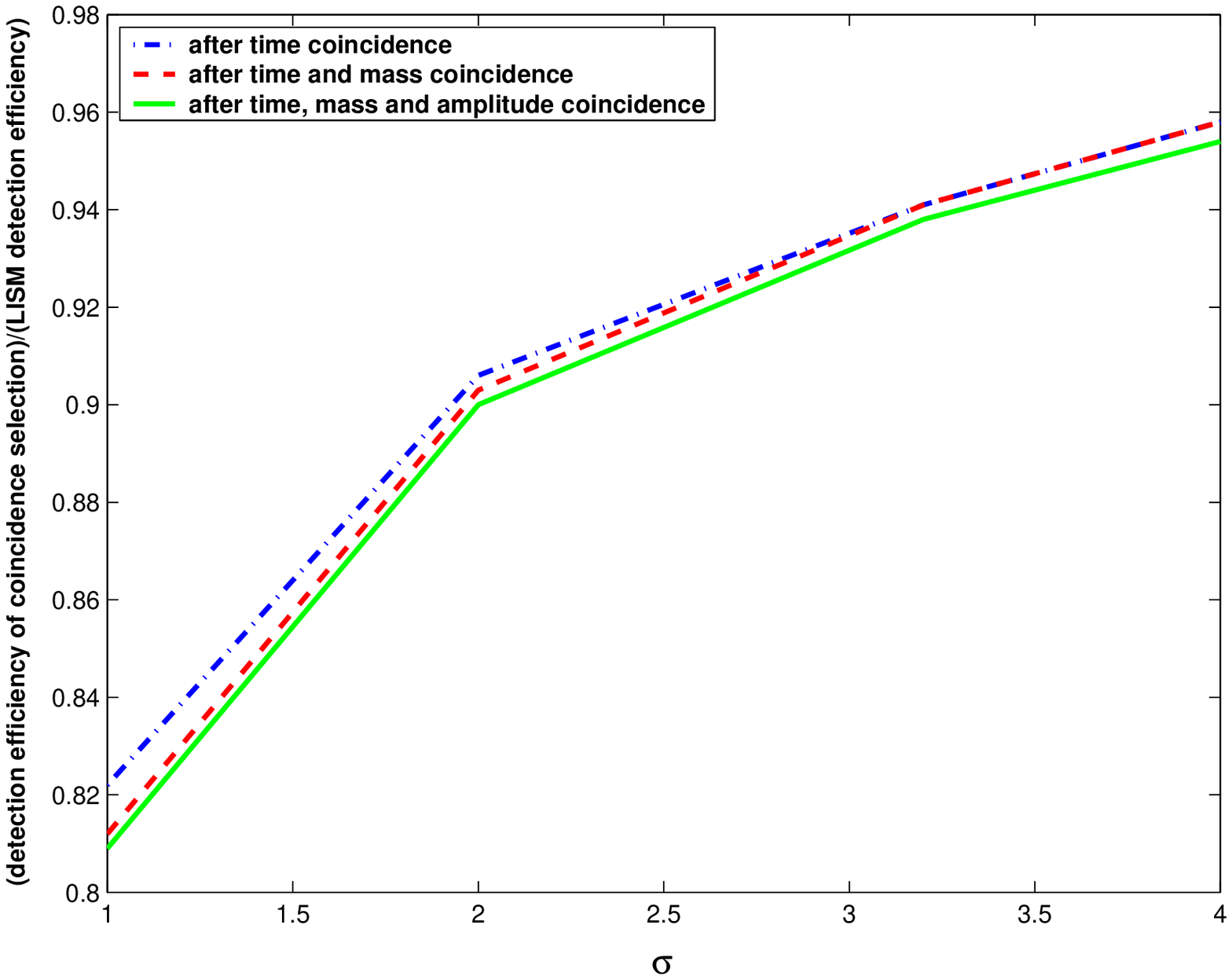}
\caption{Relative detection efficiency of the coincidence analysis compared 
to the one detector efficiency of LISM, as a function of a parameter of 
coincidence selection $\sigma$. 
Dot-dashed line is the efficiency after the time
selection, dashed line is the efficiency after the time and mass
selection, and solid line is the efficiency after the time - mass - amplitude
selection.}
\label{fig:effcoin}
\end{center}
\end{figure}

\begin{figure}[htpd]
\begin{center}
\epsfbox{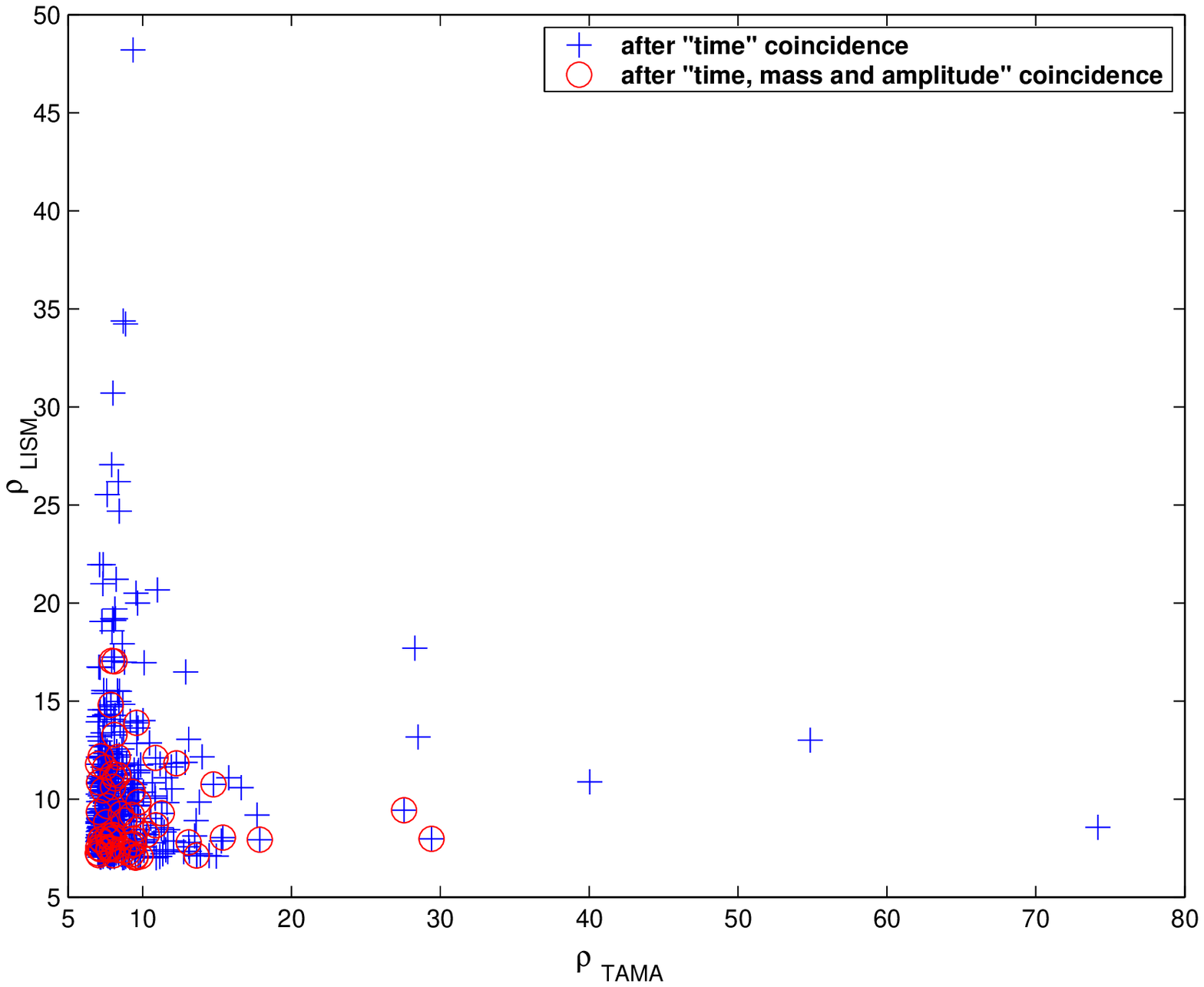}
\caption{$\rho_{\mbox{\rm\tiny TAMA}}$
 - $\rho_{\mbox{\rm\tiny LISM}}$ scatter plots. 
The + marks represent the events
survived time selection, and circle marks represent the events
survived the time, mass and amplitude selection.}
\label{fig:scatter1}
\end{center}
\end{figure}

\begin{figure}[htpd]
\begin{center}
\epsfbox{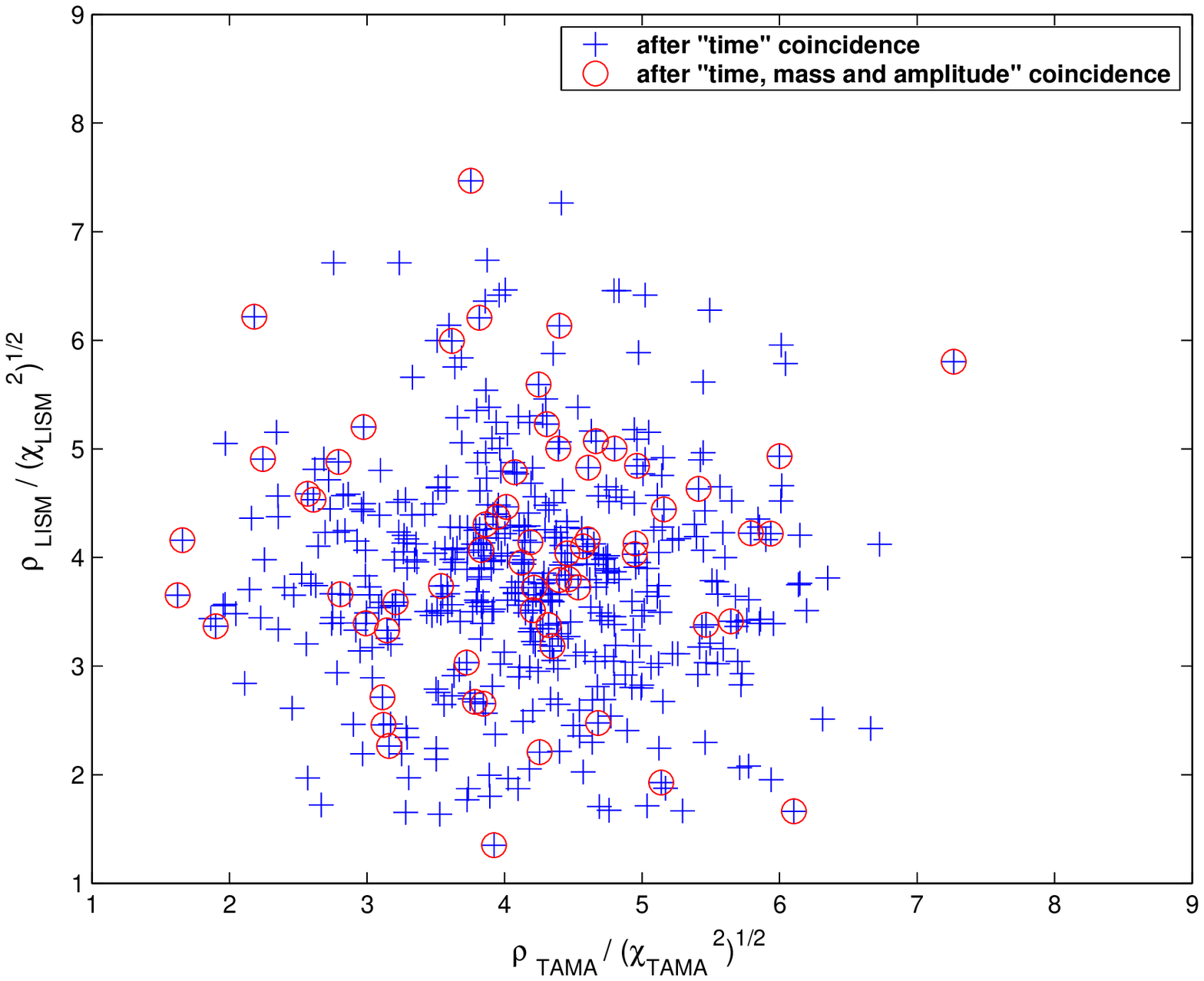}
\caption{$\rho_{\mbox{\rm\tiny TAMA}}/\sqrt{\chi_{\mbox{\rm\tiny TAMA}}^2}$ 
- $\rho_{\mbox{\rm\tiny LISM}}/\sqrt{\chi_{\mbox{\rm\tiny LISM}}^2}$ scatter plots.
The + marks represent the events
survived the time selection and circle marks represent the events
survived the time, mass and amplitude selection.}
\label{fig:scatter2}
\end{center}
\end{figure}



\section*{References} 
\begin{thebibliography}{99} 
\bibitem{tama} K.\ Tsubono, in {\it Gravitational Wave Experiments},
     edited by E.\ Coccia, G.\ Pizzella, and F.\ Ronga, (World Scientific,
     Singapore,1995).

\bibitem{ligo} A.\ Abramovici et al., Science {\bf 256}, 325 (1992).

\bibitem{geo} K.\ Danzmann et al., in {\it Gravitaional Wave Experiments},
     edited by E.\ Coccia, G.\ Pizzella, and F.\ Ronga, (World Scientific,
     Singapore,1995).

\bibitem{virgo} B.\ Caron et al., in {\it Gravitaional Wave Experiments},
     edited by E.\ Coccia, G.\ Pizzella, and F.\ Ronga, (World Scientific,
     Singapore, 1995).

\bibitem{Nicholson} D.\ Nicholson et al., Phys. Lett. A, {\bf 218} ,175 (1996). 

\bibitem{takahashi} H. Takahashi et al., in preparation.

\bibitem{Blanchet} L.\ Blanchet et al., Phys. Rev. Lett. {\bf 74}, 3515 (1995).

\bibitem{tama1} H.\ Tagoshi et al., Phys. Rev. {\bf D63}, 062001 (2001). 

\bibitem{tanaka-tagoshi} T.\ Tanaka and H.\ Tagoshi, Phys. Rev. {\bf D62},
082001 (2000).

\bibitem{Allen} B.\ Allen et al., Phys. Rev. Lett. {\bf 62}, 1489 (1999).

\bibitem{Cutler} C.\ Cutler and \'E.\ E.\ Flanagan, 
Phys. Rev. {\bf D49}, 082001 (1994).

\bibitem{Amaldi} E.\ Amaldi et al., Astron. Astrophys. {\bf 216}, 325 (1989).

\bibitem{Astone} P.\ Astone et al. , Phys. Rev. {\bf D59}, 122001 (1999).

\bibitem{particle} For example,\ Particle Data Group,\ {\it Review of particle 
Physicis}, Phys. Letts.{\bf B204}, 81 (1988).

\end{thebibliography}
\end{document}